# ENERGY LOCALIZATION AND TRANSFER IN AUTORESONANT WEAKLY DISSIPATIVE ANHARMONIC CHAINS

AGNESSA KOVALEVA*

Space Research Institute, Russian Academy of Sciences, Moscow 117997, Russia

## HIGHLIGHTS

- A weakly dissipative nonlinear chain driven by harmonic forcing with a slowly-varying frequency chain demonstrates autoresonance in a bounded time interval.
- The emergence and the duration of autoresonance depend on the interplay between the structural and excitation parameters.
- Autoresonant energy localization with energy equipartition between the autoresonant oscillators in the entire chain or in a part of the chain is observed.

*E-mail address: agnessa_kovaleva@hotmail.com

## Abstract

In this work, we develop an analytical framework to explain the influence of dissipation and detuning parameters on the emergence and stability of autoresonance in a strongly nonlinear weakly damped chain subjected to harmonic forcing with a slowly-varying frequency. Using the asymptotic procedures, we construct the evolutionary equations, which describe the behavior of the array under the condition of 1:1 resonance and then approximately compute the slow amplitudes and phases as well as the duration of autoresonance. It is shown that, in contrast to autoresonance in a non-dissipative chain with unbounded growth of energy, the energy in a weakly damped array being initially at rest is growing only in a bounded time interval up to an instant of simultaneous escape from resonance of all autoresonant oscillators. Analytical conditions of the emergence and stability of autoresonance are confirmed by numerical simulations.

### 1. Introduction and motivation

In this work, we investigate the emergence and stability of autoresonance in a strongly nonlinear chain of weakly damped oscillators driven by an external harmonic forcing with a slowly-varying frequency. The problem is investigated within the frames of the standard approach to autoresonance phenomena in nonlinear systems, which employs an intrinsic property of a nonlinear oscillator to change both its amplitude and natural frequency when the driving frequency changes. The ability of a nonlinear oscillator to stay captured into resonance due to variance of its structural or/and excitation parameters is known as *autoresonance*.

After first studies for the purpose of particles accelerations [1-3], autoresonance has become a very active field of research in different areas of natural science and engineering, from celestial mechanics [4-6] and plasmas [6-8] to mechanical structures [9, 10]. Over the years, the attention was focused on physical processes described as oscillator chains with the low number of particles. But the results of numerical simulations for a single oscillator with considerable detuning rate (e.g., see [11]) and references therein) demonstrated the upper bounding level of the forcing amplitude than the critical excitation level needed for the transition to autoresonance from zero initial state of a similar oscillator subject to forcing with slowly-varying frequency [12]. This discrepancy is attributed to different resonant properties of the oscillators. The asymptotic approach from [12] was then extended to the study of autoresonance in a weakly damped quasi-linear Duffing oscillator [13], where the autoresonant response enhancement only in a bounded time interval was demonstrated and the duration of autoresonance and the amplitude attainable were calculated for the first time.

In the next step, the procedures from [12, 13] were employed to investigate the dynamics of one-dimensional undamped quasi-linear [14] and strongly nonlinear [15] oscillator chains. The analysis of autoresonance in an $n$-particle weakly dissipative quasi-linear chain was suggested in the recent work [16]. It was shown that autoresonance in the dissipative quasi-linear chain may be observed only in a bounded time interval, in agreement with the previous simplified analysis for a single oscillator [13]. At the same time, the growth of dissipation leads to escape from resonance of either the entire chain or a part of the chain distant from the source of excitation.

The purpose of the current work is to extend the methods and the results derived in the quasi-linear theory [16] to strongly nonlinear weakly damped arrays. The main goal is to prove that autoresonance and energy localization in the weakly damped anharmonic chain driven by harmonic forcing with slowly time-varying frequency can be observed only in a bounded time interval, the length of which depends on the interaction between the structural and excitation parameters.

The rest of this article is organized as follows. In Sec. 2, we introduce the equations of autoresonant motion for a one-dimensional chain consisting of $n$ identical weakly damped Duffing oscillators with weak linear coupling between the neighbors and driven by harmonic excitation with slowly-varying frequency applied to the first oscillator. Although an anharmonic oscillator does not possess a natural frequency independent of energy of oscillations, the most effective energy transport in the nonlinear chain occurs due to 1:1 (fundamental) resonance, when the response of the chain is approximately monochromatic with the frequency close to the excitation frequency. Under this assumption, the transformations suggested, e.g., in [17], reduce the strongly nonlinear equations to the formally quasi-linear system. In analogy to the quasi-linear theory [16], the autoresonant solutions are sought with the help of the multiple time scales method [17, 18], which leads to the averaged equations for the leading-order slow envelopes.

The nonlinear averaged equations of Sec. 2 do not allow explicit analytical solutions. To clarify the physical nature of the processes, in Sec. 3 we introduce the quasi-steady equations, from which the autoresonant responses of oscillators can be approximately found, and the duration of autoresonance can be computed. In addition, escape from resonance capture due to an increase of the dissipation and/or detuning parameters is depicted. In Sec. 3, this procedure is performed for a basic single oscillator. The convergence of the exact autoresonant amplitude to the backbone curve (i.e., the frequency-amplitude dependence [19, 20]) is proved. Note that this result remains valid for multi-particle chains.

The results for a single oscillator are extended to multi-particle arrays in Sec. 4. We discuss the convergence of the autoresonant amplitudes to the backbone curve common for all autoresonant oscillators. The existence of a common backbone curve is associated with the approximate equipartition of energy between the autoresonant oscillators at large times. This implies that the study of autoresonance in the long-length arrays can be performed in the same way as for a single oscillator, that is, it can be reduced to the comparison between the exact (numerical) solutions and their quasi-steady approximations. An increase of the dissipation and/or detuning parameters leads to successive escape from resonance of all oscillators starting from the most distant from the source of energy. This effect suggests *energy localization* in the autoresonant interval of the chain adjacent to source of energy. Numerical simulations for two- and four-particle arrays confirm the analytical predictions.

Concluding remarks in Sec. 5 indicate that autoresonance can serve as an effective instrument to provide the response enhancement in the weakly damped nonlinear oscillators due to a proper choice of the excitation parameters.

Note that this work discusses only the models with a single excitation but autoresonance in the arrays with several resonant excitations can be studied in the similar way. Illustrating examples in Sec. 3, Sec. 4 demonstrate the oscillator responses for the chains with $n = 1, 2, 4$. Multi-particle arrays with $n > 4$ can be studied with the help of the similar techniques but these arrays become more sensitive to variations of the structural and excitation parameters.

Critical thresholds for the structural and excitation parameters associated with escape from the resonance domain are briefly discussed in Appendix.

## 2. Main equations

We study the dynamics of a one-dimensional weakly dissipative nonlinear chain driven by harmonic forcing with a slowly-varying frequency applied to the first oscillator. Assuming linear coupling between the oscillators, the chain dynamics is governed by the following nonlinear equations:

$$\frac{d^2U_r}{dt^2} + \chi\frac{dU_r}{dt} + \gamma U_r^3 + \kappa\left[\eta_{r,r-1}(U_r - U_{r-1}) + \eta_{r,r+1}(U_r - U_{r+1})\right] = Q_r \sin\theta, \qquad (1)$$

$$\frac{d\theta}{dt} = \omega + \zeta(t);\ \zeta(t) = kt,$$

In (1), the variable $U_r$ denotes the absolute displacement of the *r-th* oscillator from its rest state, $r \in [1, n]$; $\gamma$ is the cubic stiffness coefficient; $\kappa$ denotes stiffness of linear coupling; $\chi$ is the coefficient of dissipation; all parameters are reduced to the unit mass. The coefficients $\eta_{r,k} = \{1,\ k\in[1,\ n];\ 0,\ k = 0,\ k = n + 1\}$ indicate that the end oscillators are unilaterally coupled with the neighboring elements. Since external forcing is applied to the first oscillator, we let $Q_1 = Q$, $Q_r = 0$ at $r \geq 2$. The chain is initially at rest, i.e. $U_r = 0$, $V_r = dU_r/dt = 0$ at $t = 0$. Recall that zero initial conditions determine the so-called Limiting Phase Trajectory (LPT) corresponding to maximum possible energy transfer from the source of energy to the receiver [12, 21]. This definition remains valid for the multi-particle arrays, where the response of each oscillator is considered as a source of energy for its neighbors.

For further analysis, it is convenient to reduce (1) to the dimensionless form. First, the dimensionless stiffness of weak coupling $\varepsilon = \kappa/(2\omega^2) << 1$ is considered as a small parameter of the problem. In the next step, weak dissipation and low-level external forcing with small detuning rate are rescaled as follows:

$$\alpha = 3\gamma/4\omega^2,\ 2\varepsilon f = \alpha^{1/2}A/\omega^2,\ 2\varepsilon\delta = \chi/\omega;\ \varepsilon^2\beta = k/\omega^2.$$

Finally, we define the dimensionless space variables $u_r = \alpha^{1/2}U_r$ and the fast- and slow-time scales $\tau_0 = \omega t$ and $\tau = \varepsilon\tau_0$, respectively. Substituting the new variables into (1), we obtain the following dimensionless equations:

$$\frac{d^2u_r}{d\tau_0^2} + 2\varepsilon\delta\frac{du_r}{d\tau_0} + \frac{4}{3}u_r^3 + 2\varepsilon\left[\eta_{r,r-1}(u_r - u_{r-1}) + \eta_{r,r+1}(u_r - u_{r+1})\right] = 2\varepsilon f_r \sin\theta, \quad (2)$$

$$\frac{d\theta}{d\tau_0} = 1 + \varepsilon\zeta_0(\tau),\ \zeta_0(\tau) = \beta\tau.$$

where $f_1 = f > 0$ but $f_r = 0$ at $r \in [2, n]$. Although the generating nonlinear system $\frac{d^2u_r}{d\tau_0^2} + \frac{4}{3}u_r^3 = 0$ does not possess a spectrum independent of energy of oscillations, intense energy transport is studied under the assumption of 1:1 (fundamental) resonance, i.e., under the condition that the response of each oscillator in the chain has a dominant harmonic component with the frequency close to the excitation frequency (e.g., see [20]). Under this assumption, the equations of motion are rewritten in the form suggested in [15]:

$$\frac{d^2u_r}{d\tau_0^2} + 2\varepsilon\delta\frac{du_r}{d\tau_0} + u_r + 2\varepsilon[\sigma\left(\frac{4}{3}u_r^3 - u_r\right) + \eta_{r,r-1}(u_r - u_{r-1}) + \eta_{r,r+1}(u_r - u_{r+1})] = \varepsilon f_r \sin\theta$$

$$\frac{d\theta}{d\tau_0} = 1 + \varepsilon\zeta_0(\tau),\ r \in [1, n], \quad (3)$$

where $2\varepsilon\sigma = 1$. Since system (3) is formally quasi-linear, the asymptotic transformations from [15] can be employed. First, we introduce the new slow-fast variables:

$$\psi_r = \left(\frac{du_r}{d\tau_0} + iu_r\right)e^{-i\theta}\,,\ \psi_r^* = \left(\frac{du_r}{d\tau_0} - iu_r\right)e^{i\theta}\,. \tag{4}$$

It follows from (4) that the real-valued amplitudes and phases of oscillations take the form $A_r = |\psi_r|$, $\Theta_r = \arg\psi_r$. From (3), (4), we derive the following (still exact) equations in the standard form for the envelopes $\psi_r$:

$$\frac{\mathrm{d}\psi_r}{\mathrm{d}\tau_0} = -\varepsilon\delta\psi_r +$$

$$i\varepsilon\left[\sigma\left(|\psi_r|^2 - 1\right)\psi_r - \zeta_0(\tau)\psi_r + \eta_{r,r-1}(\psi_r - \psi_{r-1}) + \eta_{r,r+1}(\psi_r - \psi_{r+1}) - f_r + G_r(\psi, \psi^*, \vartheta)\right]$$

$$\frac{d\theta}{d\tau_0} = 1 + \varepsilon\zeta_0(\tau) \tag{5}$$

and similar equations for the complex-conjugate envelopes $\psi_r^*$, $r \in [1, n]$. Since the system is initially at rest, then $\psi_r = \psi_r^* = 0$ at $\tau = 0$. The term $G_r(\psi, \psi^*, \theta)$ generated by the nonlinearity $u_r^3$ can be presented as the sum of zero-mean higher harmonics in $\theta$ with the coefficients depending on $\psi_r$, $\psi_r^*$ [15].

As in the previous works, we employ the method of multiple scales [18] to approximately compute solutions of (5). Taking into account the dependence of the right-hand side of (5) on the parameter $\sigma = 1/(2\varepsilon)$, we consider the following asymptotic representations of the solution:

$$\psi_r(\tau_0, \tau, \varepsilon) = \psi_r^{(0)}(\tau, \varepsilon) + \varepsilon\psi_r^{(2)}(\tau_0, \tau, \varepsilon) + O(\varepsilon^2),\ r \in [1, n]. \tag{6}$$

The asymptotic expansion provides the following equations for the leading-order slow components $\psi_r^{(0)}(\tau, \varepsilon)$:

$$\frac{\mathrm{d}\psi_r^{(0)}}{\mathrm{d}\tau} = -\delta\,\psi_r^{(0)} +$$

$$+i\left[\sigma\left(|\psi_r^{(0)}|^2 - 1\right)\psi_r^{(0)} - \zeta_0(\tau)\psi_r^{(0)} + \eta_{r,r-1}\left(\psi_r^{(0)} - \psi_{r-1}^{(0)}\right) + \eta_{r,r+1}\left(\psi_r^{(0)} - \psi_{r+1}^{(0)}\right) - f_r\right] \tag{7}$$

with zero initial conditions $\psi_r^{(0)} = 0$ at $\tau = 0$. Finally, the change of variables

$$\psi_r^{(0)} = a_r e^{i\Delta_r}, a_r = \left|\psi_r^{(0)}\right|,\ \Delta_r = \arg\psi_r^{(0)} \tag{8}$$

transforms (7) into the following equations for the real amplitudes $a_r(\tau,\varepsilon)$ and phases $\Delta_r(\tau,\varepsilon)$:

$$\frac{da_r}{d\tau} = -\delta a_r + [\eta_{r,r-1}a_{r-1}\sin(\Delta_{r-1} - \Delta_r) + \eta_{r,r+1}a_{r+1}\sin(\Delta_{r+1} - \Delta_r)] - f_r \sin\Delta_r, \tag{9}$$

$$a_r\frac{d\Delta_r}{d\tau} = \sigma(a_r^2 - 1)a_r - \zeta_0(\tau)a_r +$$

$$+[\eta_{r,r-1}(a_r - a_{r-1}\cos(\Delta_{r-1} - \Delta_r)) + \eta_{r,r+1}(a_r - a_{r+1}\cos(\Delta_{r+1} - \Delta_r))] - f_r\cos\Delta_r,$$

with zero amplitudes and uncertain phases at $\tau = 0$. To overcome this uncertainty, one needs to solve (7) with zero initial conditions and then calculate the real-valued amplitudes and phases by formulas (8). It is important to note that, due to the convergence

$$\left|\left|\psi_r(\theta,\tau,\varepsilon)\right| - \left|\psi_r^{(0)}(\tau,\varepsilon)\right|\right| \to 0 \text{ as } \varepsilon \to 0, \tag{10}$$

valid for dissipative systems in an infinite time interval [22], the slow dynamics of the system depicts main physical features of the original process.

For brevity, the explicit dependence of the slow variables on the parameter $\varepsilon$ will be omitted from further consideration; the solutions $a_r(\tau,\varepsilon)$, $\Delta_r(\tau,\varepsilon)$ and their derivatives will be expressed as $a_r(\tau)$, $\Delta_r(\tau)$, etc.

## 3. Autoresonance in a single oscillator

It was recently demonstrated [16] that autoresonant amplitudes in the quasi-linear chain can be depicted as the superposition of fast oscillations on the adiabatically varying backbone curves. We show that this result remains valid for the anharmonic system. To this end, we describe the solutions of (9) as follows:

$$a_r(\tau) = \tilde{a}_r(\tau) + \rho_r(\tau), \Delta_r(\tau) = \tilde{\Delta}_r(\tau) + \varphi_r(\tau)\ , \tag{11}$$

where the terms $\tilde{a}_r, \tilde{\Delta}_r$ represent the quasi-steady approximations satisfying the equations:

$$P_r = \frac{d\tilde{a}_r}{d\tau} = 0, Q_r = \frac{d\tilde{\Delta}_r}{d\tau} = 0, r \in [1, n], \tag{12}$$

the terms $\rho_r$, $\varphi_r$ describe the deviations of the exact solutions from their quasi-steady states. Further analysis will prove small contributions of the terms $\rho_r(\tau)$, $\varphi_r(\tau)$ into expressions (11).

*3.1. Autoresonance response of a single oscillator*

In order to clarify and justify the application of the asymptotic procedures, we begin with an example of a single oscillator. We recall that autoresonance in the forced oscillator can be considered as a necessary condition of autoresonance in the passive attachment [14-16]. The assumptions $a_1 \sim O(1)$, $a_r \sim o(1)$, $r \geq 2$, lead to the following equations for the excited oscillator:

$$\frac{d^2u_1}{d\tau_0^2} + 2\varepsilon\delta\frac{du_1}{d\tau_0} + u_1 + 2\varepsilon\sigma\left(\frac{4}{3}u_1^3 - u_1\right) = 2\varepsilon f \sin\theta(t), \tag{13}$$

$$\frac{d\theta}{d\tau_0} = 1 + \varepsilon\zeta_0(\tau), \zeta_0(\tau) = \beta\tau,$$

with initial conditions $u_1(0) = v_1(0) = 0$, $\theta(0) = 0$ corresponding to the Limiting Phase Trajectory of the single oscillator [12]. It follows from (9) that the averaged system for the single oscillator at $r = 1$ takes the form:

$$\frac{da_1}{d\tau} = -\delta a_1 - f\sin\Delta_1, \tag{14}$$

$$a_1\frac{d\Delta_1}{d\tau} = \sigma(a_1^2 - 1)a_1 - \zeta_0(\tau)a_1 - f\cos\Delta_1,$$

with initial conditions $a_1(0) = 0$, $\Delta_1 = -\pi/2$. The quasi-steady states of (14) can be found from the following equations:

$$\sin\tilde{\Delta}_1 = -\frac{\delta}{f}\tilde{a}_1, \tag{15}$$

$$\sigma(\tilde{a}_1^2 - 1) - \zeta_0(\tau) - \frac{f}{\tilde{a}_1}\cos\tilde{\Delta}_1 = 0,$$

where $\sigma = 1/(2\varepsilon)$. The approximate solutions of (15) are sought as follows:

$$\tilde{a}_1(\tau) = \bar{a}(\tau) + \varepsilon\alpha_1(\tau),$$

$$\sin\tilde{\Delta}_1(\tau) = -\frac{\delta}{f}\left(\bar{a}(\tau) + \varepsilon\alpha_1(\tau)\right) = \sin\bar{\Delta}_1(\tau) - \varepsilon\frac{\delta}{f}\alpha_1(\tau), \tag{16}$$

$$\sin \bar{\Delta}_1(\tau) = -\frac{\delta}{f}\bar{a}(\tau),$$

where the main term

$$\bar{a}(\tau) = \sqrt{1 + 2\varepsilon\zeta_0(\tau)} \tag{17}$$

plays the role of the backbone curve independent of the forcing amplitude. Substituting (16), (17) into (15) and considering the asymptotic expansion as $\varepsilon \to 0$, we obtain

$$\begin{aligned} \alpha_1(\tau) &= \frac{f\cos\bar{\Delta}_1(\tau)}{\bar{a}^2(\tau)} + \varepsilon O\left(\frac{1}{\bar{a}^5(\tau)}\right), \\ \tilde{a}_1(\tau) &= \bar{a}(\tau) + \varepsilon\frac{f\cos\bar{\Delta}_1(\tau)}{\bar{a}^2(\tau)} + \varepsilon^2 O\left(\frac{1}{\bar{a}^5(\tau)}\right), \\ \sin\tilde{\Delta}_1(\tau) &= -\frac{\delta}{f}\left[\bar{a}(\tau) + \varepsilon\frac{f\cos\bar{\Delta}_1(\tau)}{\bar{a}^2(\tau)}\right] + \varepsilon^2 O\left(\frac{1}{\bar{a}^5(\tau)}\right). \end{aligned} \tag{18}$$

Expressions (18) imply that

$$\tilde{a}_1(\tau) \to \bar{a}(\tau),\ \sin\tilde{\Delta}_1(\tau) \to \sin\bar{\Delta}_1(\tau) \tag{19}$$

at large time. It follows from (18), (19) that the instant $\tau = \tau_{AR}$ of escape from resonance and the values $\bar{a}(\tau_{AR})$, $\bar{\Delta}_1(\tau_{AR})$ can be found from the following equations:

$$\sin\bar{\Delta}_1(\tau_{AR}) = -\delta\frac{\bar{a}(\tau_{AR})}{f} = -1,\ \bar{a}(\tau_{AR}) = \frac{f}{\delta}, \tag{20}$$

$$\tau_{AR} = \frac{1}{2\varepsilon\beta}\left[\left(\frac{f}{\delta}\right)^2 - 1\right].$$

Equations (18) - (20) approximately describe autoresonance in the oscillator with dissipation $\delta < \delta_*^{(2)} = f$. But it follows from (18), (20) that autoresonance may occur only if $\delta << \delta_*^{(2)}$. This means that the critical parameter $\delta_*^{(2)}$ does not depict an exact boundary of the resonance domain but indicate a threshold which cannot be exceeded by the damping coefficient $\delta$ in the resonance regime. At the same time, the computed parameters (20) agree with the numerical results in the autoresonant oscillator (Fig. 1) even if the formal boundary of the admissible domain is unidentified. Similar effects for quasi-linear arrays have been discussed in [16].

*3.2. Deviations $\rho_1(\tau)$, $\varphi_1(\tau)$*

The equations for the deviations $\rho_1(\tau)$, $\varphi_1(\tau)$ can be derived from (14), (15). As in the previous work [16], the deviations are assumed to be small at relatively large times (see Figs. 1, 2). Under this assumption, the equations for small deviations near the terminal point $\tau = \tau_{AR}$ take the form

$$\frac{d\rho_1}{d\tau} + \delta\rho_1 = -\frac{d\bar{a}}{d\tau} \tag{21}$$

$$\frac{d\varphi_1}{d\tau} + \frac{f}{\bar{a}}\varphi_1 = 2\varepsilon^{-1}\bar{a}\rho_1,$$

The notation $\rho_1 = \varepsilon\rho$, $\varphi_1 = \varphi$ reduce (21) to the form:

$$\frac{d\rho}{d\tau} + \delta\rho = -\frac{\beta}{\sqrt{1+2\varepsilon\beta\tau}}\ , \tag{22}$$

$$\frac{d\varphi}{d\tau} + \frac{f}{\sqrt{1+2\varepsilon\beta\tau}}\varphi = 2\rho\sqrt{1+2\varepsilon\beta\tau}.$$

The solutions of (22) are expressed as

$$\rho(\tau) = C_1 e^{-\delta(\tau-T)} - \beta e^{-\delta\tau}\int_T^\tau \frac{e^{\delta s}}{\sqrt{1+2\varepsilon\beta s}}ds, \tag{23}$$

$$\varphi(\tau) = C_2 e^{-[\Phi(\tau)-\Phi(T)]} + 2e^{-\Phi(\tau)}\int_T^\tau e^{\Phi(s)}\rho(s)\sqrt{1+2\varepsilon\beta s}ds,$$

with constants $C_{1,2}$ determined through the initial values $\rho(T)$ and $\varphi(T)$, respectively; the instant $T$ is chosen not far from the terminal point $\tau_{AR}$ to ensure moderate deviations of $\bar{\Delta}_1(\tau)$ from the terminal value. The exponent $\Phi(\tau)$ is given by

$$\Phi(\tau) = f\int_0^\tau \frac{ds}{\sqrt{1+2\varepsilon\beta s}} = \frac{f}{\varepsilon\beta}[\sqrt{1+2\varepsilon\beta\tau} - 1] \to f\tau \text{ as } \varepsilon \to 0.$$

Using the mean value theorem [23], we get

$$I_1(\tau) = \int_T^\tau \frac{e^{\delta s}}{\sqrt{1+2\varepsilon\beta s}}ds = \frac{e^{\delta\tau_1}}{\sqrt{1+2\varepsilon\beta\tau_1}}(\tau-\tau_1), T < \tau_1 < \tau, \tag{24}$$

$$I_2(\tau) = \int_T^\tau e^{\Phi(s)}\rho(s)\sqrt{1+2\varepsilon\beta s}ds = e^{\Phi(\tau_2)}\rho(\tau_2)\sqrt{1+2\varepsilon\beta\tau_2}(\tau-\tau_2), T < \tau_2 < \tau.$$

Expressions (23), (24) imply the convergence $\rho(\tau) \to 0$, $\varphi(\tau) \to 0$ as $\tau \to \infty$ and, therefore, the convergence

$$a_1(\tau) \to \bar{a}(\tau), \Delta_1(\tau) \to \bar{\Delta}_1(\tau) \tag{25}$$

at large times. This means that the quasi-steady terms $\bar{a}(\tau)$, $\bar{\Delta}_1(\tau)$ can be considered as the main approximations to the exact solutions of (14) at large times. One can easily check that similar conditions hold for multi-particle arrays. This implies that the analysis of a multi-particle array can be reduced to the comparison between the exact (numerical) slow variables and their analytical quasi-steady approximations.

*3.3. Numerical results*

Figure 1 depicts autoresonance in the oscillator with parameters $\varepsilon = 0.13$, $f = 0.7$, $\beta = 0.04$, $\delta = 0.04$.

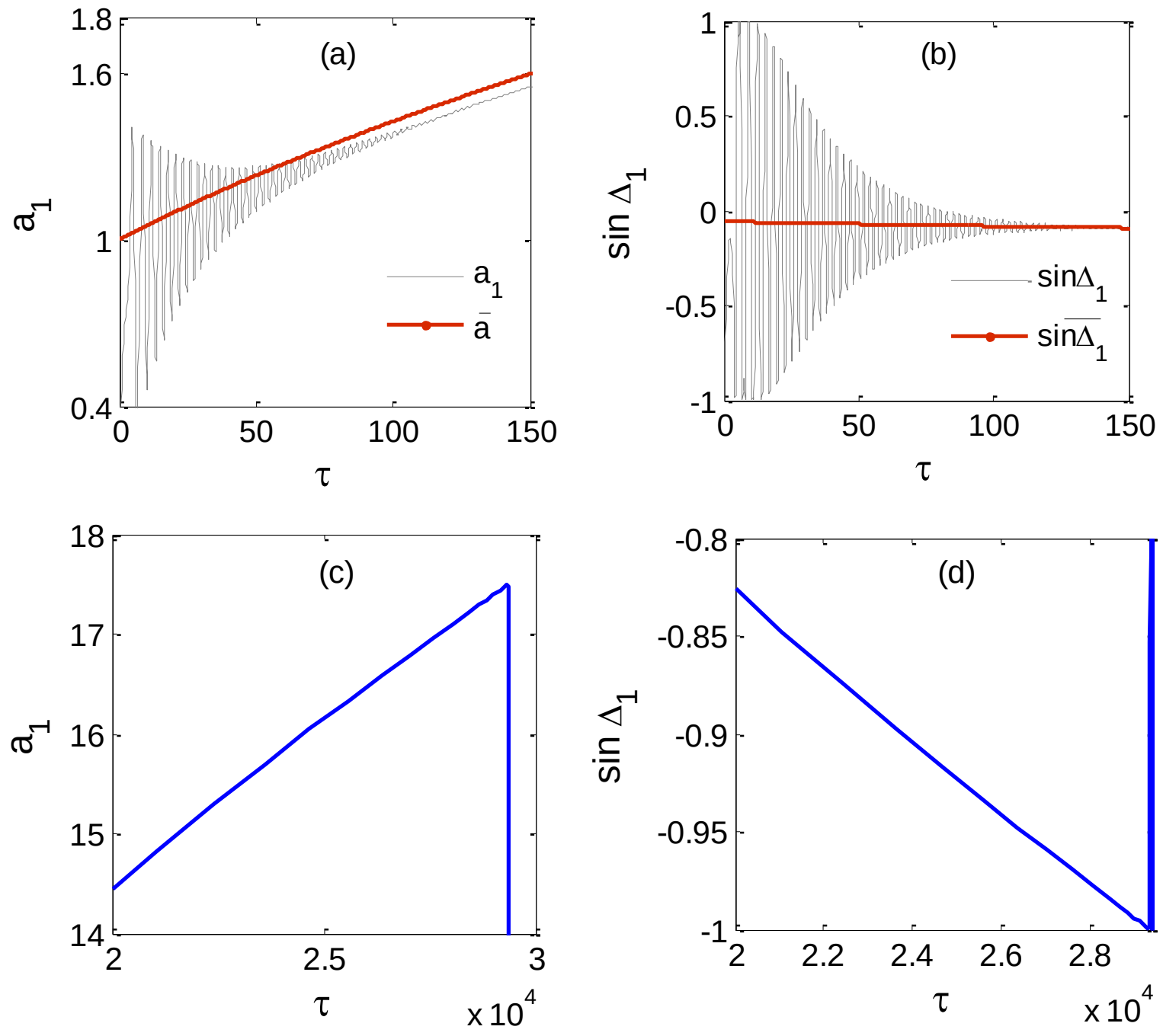

**Fig. 1.** Autoresonance in oscillator (14) with parameters $\varepsilon = 0.13$, $f = 0.7$, $\beta = 0.04$, $\delta = 0.04$: the amplitude $a_1(\tau)$ (plot (*a*)) and the corresponding function $\sin\Delta_1(\tau)$ (plot (*b*)) in the initial time interval; the solid lines depicts the main approximations $\bar{a}(\tau)$ and $\sin\bar{\Delta}_1(\tau)$, respectively. Escape from resonance capture are illustrated in plots (*c*) and (*d*), respectively.

It follows from (A.2), (A.3) and Fig. 10 that the parameters $(\varepsilon, f) \in D_0$, that is, the undamped oscillator is captured into resonance. This implies that oscillator (14) with the chosen parameters exhibits autoresonance provided that $\beta << \beta^*$ and $\delta << \delta_*^{(2)} = 0.7$. The critical thresholds $\beta^* = 0.058$ for this oscillator is established in Appendix.

We obtain from (20) that $\tau_{AR} = 29350$; $\bar{a}(\tau_{AR}) = 17.5$. Figures 1(*a*),1(*b*) illustrate the fast convergence of the exact solutions to their quasi-steady approximations in the initial interval of motion; escapes from resonance capture at $\tau \approx \tau_{AR}$ are depicted in Figs. 1(*c*), 1(*d*). It can be easily checked that amplitude $a_1(\tau_{AR}) \approx \bar{a}(\tau_{AR})$; the differences between the instant of escape from resonance and the amplitude attainable and their quasi-steady approximations become indistinguishable in the chosen time- and space-scales. This agreement underlines the role of the backbone curve $\bar{a}(\tau)$ in the construction of the approximate solutions.

Figure 2 demonstrates exit from autoresonance due to the growth of the dissipation and/or detuning coefficients.

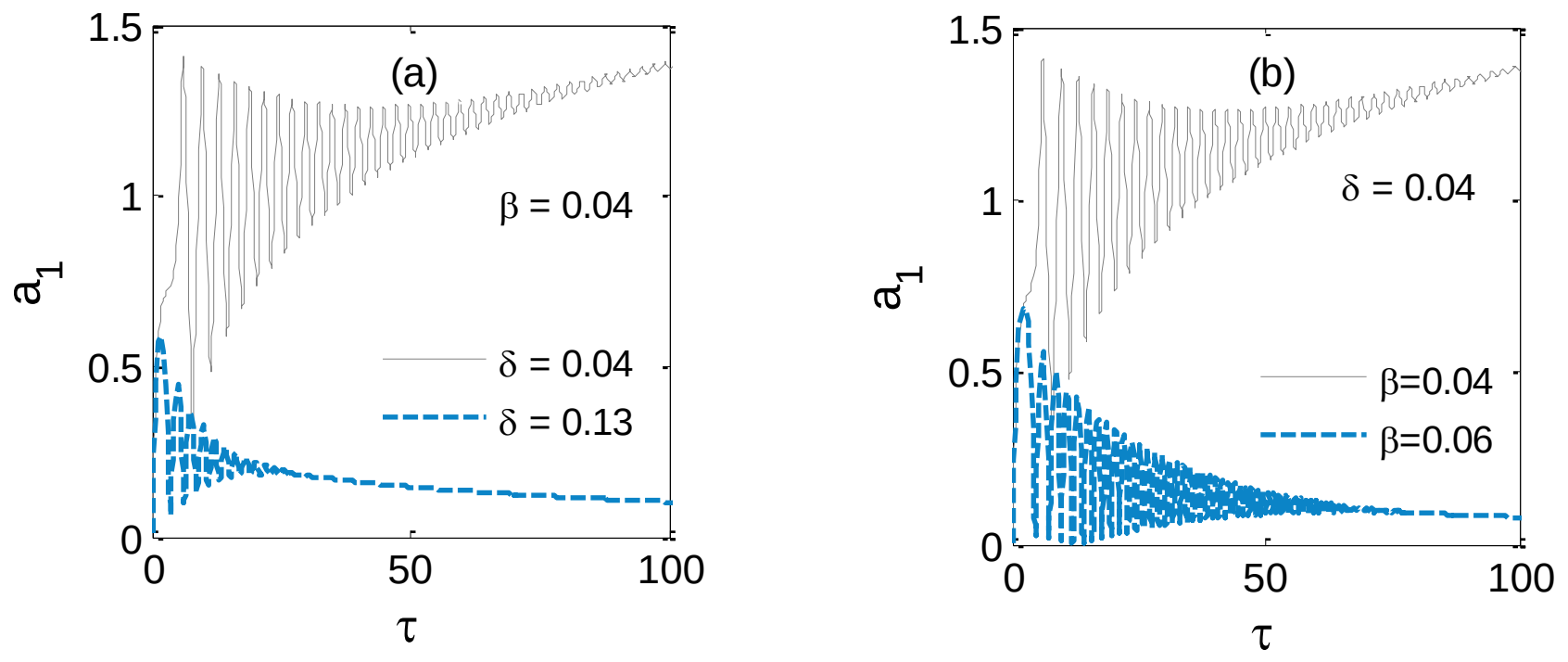

**Fig. 2.** Autoresonant and non-resonant oscillations of oscillator (14) with parameters ($\varepsilon = 0.13$, $f = 0.7$) $\in D_0$ and different dissipation and detuning coefficients.

## 4. Autoresonance in multi-particle chains

In this section, we investigate autoresonant energy transfer and localization in the multi-particle chain. Due to similarity of the response of the autoresonant oscillators to the response of the single oscillator, the analysis of the array is performed in the same way as in Sec. 3.

Expressions (12) for the *n*-particle array take the form:

$$P_r = -\delta a_r + [\eta_{r,r-1} a_{r-1} \sin(\Delta_{r-1} - \Delta_r) + \eta_{r,r+1} a_{r+1} \sin(\Delta_{r+1} - \Delta_r)] - f_r \sin \Delta_r = 0 \quad (26)$$

$Q_r = \sigma(a_r^2 - 1) - \zeta_0(\tau) +$

$+[\eta_{r,r-1}(1 - \frac{a_{r-1}}{a_r}\cos(\varDelta_{r-1} - \varDelta_r)) + \eta_{r,r+1}(1 - \frac{a_{r+1}}{a_r}\cos(\varDelta_{r+1} - \varDelta_r))] - \frac{f_r}{a_r}\cos\varDelta_r,$

where the forcing amplitude $f_1 = f$ but $f_r = 0$ at $r \in [2, n]$. The autoresonant solutions are sought in the form (11), namely, $a_r(\tau) = \tilde{a}_r(\tau) + \rho_r(\tau), \varDelta_r(\tau) = \widetilde{\varDelta}_r(\tau) + \varphi_r(\tau)$, where

$$\tilde{a}_r(\tau) = \bar{a}(\tau) + \varepsilon\alpha_r(\tau), \tag{27}$$

$$\alpha_1(\tau) = \frac{f\cos\bar{\varDelta}_1(\tau)}{\bar{a}^2(\tau)} + \varepsilon O(\frac{1}{\bar{a}^5(\tau)}), \alpha_r(\tau) = \varepsilon O(\frac{1}{\bar{a}^5(\tau)}).$$

As in the previous section, expressions (27) can be employed to demonstrate the convergence of the slow amplitudes to the backbone curve:

$$a_r(\tau) \to \bar{a}(\tau) = \sqrt{1 + 2\varepsilon\zeta_0(\tau)} \text{ as } \tau \to \infty, r \in [1, n]. \tag{28}$$

It follows from (27), (28) that the energy initially placed in the first oscillator tends to equipartition among all autoresonant oscillators at large times but the energy of the excited oscillator exceeds the mean energy of the attached oscillators in the initial time interval (see examples below). Substituting (28) into (26), we find that the energy equipartition (28) does not entail the equal phases; the quasi-steady phases $\bar{\varDelta}_r(\tau)$ obey the following equations:

$$[-\delta + \sin(\bar{\varDelta}_2 - \bar{\varDelta}_1)]\bar{a}(\tau) = f\sin\bar{\varDelta}_1(\tau),$$

$$-\delta + [\sin(\bar{\varDelta}_{r-1} - \bar{\varDelta}_r) + \sin(\bar{\varDelta}_{r+1} - \bar{\varDelta}_r)] = 0, r \in [2, n-1],$$

$$-\delta + \sin(\bar{\varDelta}_{n-1} - \bar{\varDelta}_n) = 0\ ,$$

or

$$\sin\bar{\varDelta}_1 = -\delta n\frac{\bar{a}(\tau)}{f}, \tag{29}$$

$$\sin(\bar{\varDelta}_r - \bar{\varDelta}_{r-1}) = -\delta[n - (r-1)], \ldots, \sin(\bar{\varDelta}_n - \bar{\varDelta}_{n-1}) = -\delta.$$

It follows from (29) that if $\delta \ll \delta_1^{(n)} = \frac{f}{n}$, $\delta \le \delta_r^{(n)} = \frac{1}{n-(r-1)}$, $r \in [2, n]$, or

$$\delta << \delta_*^{(n)} = \min(\frac{f}{n}, \frac{1}{n-1}), \tag{30}$$

then $\sin\bar{\varDelta}_1(\tau_{AR}) = -1$ at the instant $\tau = \tau_{AR}$ such that

$$\sin\overline{\Delta}_1(\tau_{AR}) = -\delta n\frac{\bar{a}(\tau_{AR})}{f} = -1, \bar{a}(\tau_{AR}) = \frac{f}{\delta n}, \tau_{AR} = \frac{1}{2\varepsilon\beta}[(\frac{f}{\delta n})^2 - 1] \tag{31}$$

but $|\sin(\overline{\Delta}_r - \overline{\Delta}_{r-1})| \le 1$ at $\tau \in [0, \tau_{AR}]$. Conditions (28), (31) depict energy equipartition between the autoresonant oscillators in the entire chain at large times. The convergence of the exact (numerical) amplitudes $a_r(\tau)$ to the common backbone curve $\bar{a}(\tau)$ is confirmed by numerical simulations (see below).

Expression (30) shows that critical dissipation $\delta_*^{(n)}$ drastically decreases with the growth of the chain length *n*. This means that condition (30) does not hold for the long-length arrays with relatively strong dissipation. If only $p < n$ oscillators are captured into resonance, then the autoresonant amplitudes also converge to the common backbone curve $a_r(\tau) \to \bar{a}(\tau) = \sqrt{1 + 2\varepsilon\zeta_0(\tau)}$, but the quasi-steady phases obey the following equations:

$$\sin\overline{\Delta}_1 = -\delta p\frac{\rho_\varepsilon(\tau)}{f}, \ \sin(\overline{\Delta}_r - \overline{\Delta}_{r-1}) = -\delta[p - (r-1)], r \in [2, p]. \tag{32}$$

The distant oscillators at $r \ge p + 1$ perform small non-resonant oscillations (see examples below). Under these conditions, the number *p* of the autoresonant oscillators define the so-called *localization length*.

It follows from (32) that if $\delta \ll \delta_1^{(p)} = \frac{f}{p}$, $\delta \ll \delta_r^{(p)} = \frac{1}{p-(r-1)}$, $r \in [2, p]$, or

$$\delta << \delta_*^{(p)} = \min(\frac{f}{p}, \frac{1}{p-1}) \tag{33}$$

then $\sin\overline{\Delta}_1(\tau_{AR}) = -1$ and $|\sin(\overline{\Delta}_r(\tau) - \overline{\Delta}_{r-1}(\tau))| \ll 1$ at $\tau \in [0, \bar{\tau}_{AR}]$, $r \in [2, p]$. The parameters $\tau_{AR}$ and $\bar{a}(\tau_{AR})$ for the truncated interval are defined as:

$$\sin\overline{\Delta}_1(\tau_{AR}) = -\delta p\frac{\bar{a}(\tau_{AR})}{f} = -1, \ \tau_{AR} = \frac{1}{2\varepsilon\beta}[(\frac{f}{p\delta})^2 - 1], \tag{34}$$

$$\bar{a}(\tau_{AR}) = \frac{f}{p\delta}, \ r \in [1, p].$$

### *4.1. Two-particle arrays*

In this section, we demonstrate that conditions (30), (33) as well as inequality (A.7) describe the upper limitations of the dissipation and detuning parameters, which cannot be exceeded in the autoresonance regime. First, we consider the oscillators with parameters ($\varepsilon$ = 0.13, $f$ = 0.7) $\in D_0$. Figure 3 illustrates autoresonance in both oscillators with coefficients $\delta$ = 0.04 << $\delta_*^{(2)}$ = $f/2$ = 0.35, $\beta$ = 0.04 < $\beta^*$ ≈ 0.058. Figure 3($a$) depicts the amplitudes $a_1(\tau)$, $a_2(\tau)$ at the initial interval of autoresonance. The lower solid line in Fig. 3($a$) corresponds to the backbone curve (28), while the upper dash-dotted line considers the correction terms (27) for the amplitude $a_1(\tau)$. Note that a similar effect was also revealed for a quasi-linear chain [16]. Figure 3($b$) confirms that the decaying corrections (27) drastically diminish becoming unimportant, and analytical approximations (31) become valid for both amplitudes at large times, i.e., both oscillators escape from resonance at $\tau \approx \tau_{AR}$ = 7245 with amplitudes $a_r(\tau_{AR}) \approx \bar{a}(\tau_{AR})$ = 8.75.

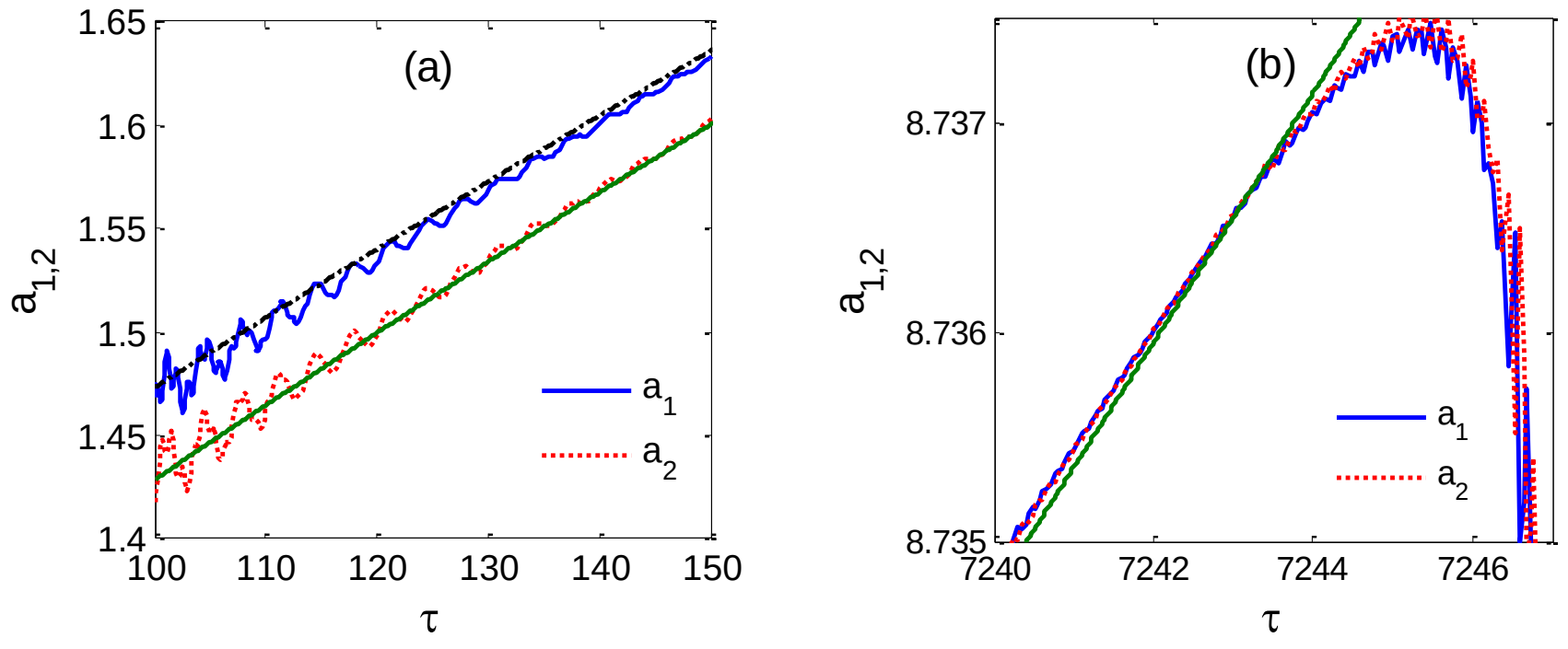

**Fig. 3.** Autoresonance in the coupled oscillators with parameters $\varepsilon$ = 0.13, $f$ = 0.7, $\beta$ = 0.04, $\delta$ = 0.04: ($a$) the amplitudes $a_1(\tau)$, $a_2(\tau)$ in the initial interval of motion; ($b$) the amplitudes $a_1(\tau)$, $a_2(\tau)$ in the final interval of motion. The solid lines in plots ($a$), ($b$) depict the backbone curve $\bar{a}(\tau)$.

The responses of the arrays with the same small rate $\beta$ = 0.04 but with the dissipation coefficients $\delta$ = 0.15 < $\delta_*^{(2)}$ and $\delta$ = 0.4 > $\delta_*^{(2)}$ are shown in Figs. 4($a$) and 4($b$), respectively. Note the coefficient $\delta$ = 0.15 is not small enough and thus, it cannot guarantee autoresonance in both particles. Figure 4($a$) illustrates autoresonance in the first oscillator and non-resonant oscillations of the second oscillator at $\delta = 0.15$; the instant of escape from resonance of the first oscillator is close to $\tau_{AR} \approx 1998$. Figure 4(*b*) depicts the non-resonant amplitudes of both oscillators at $\delta = 0.4$.

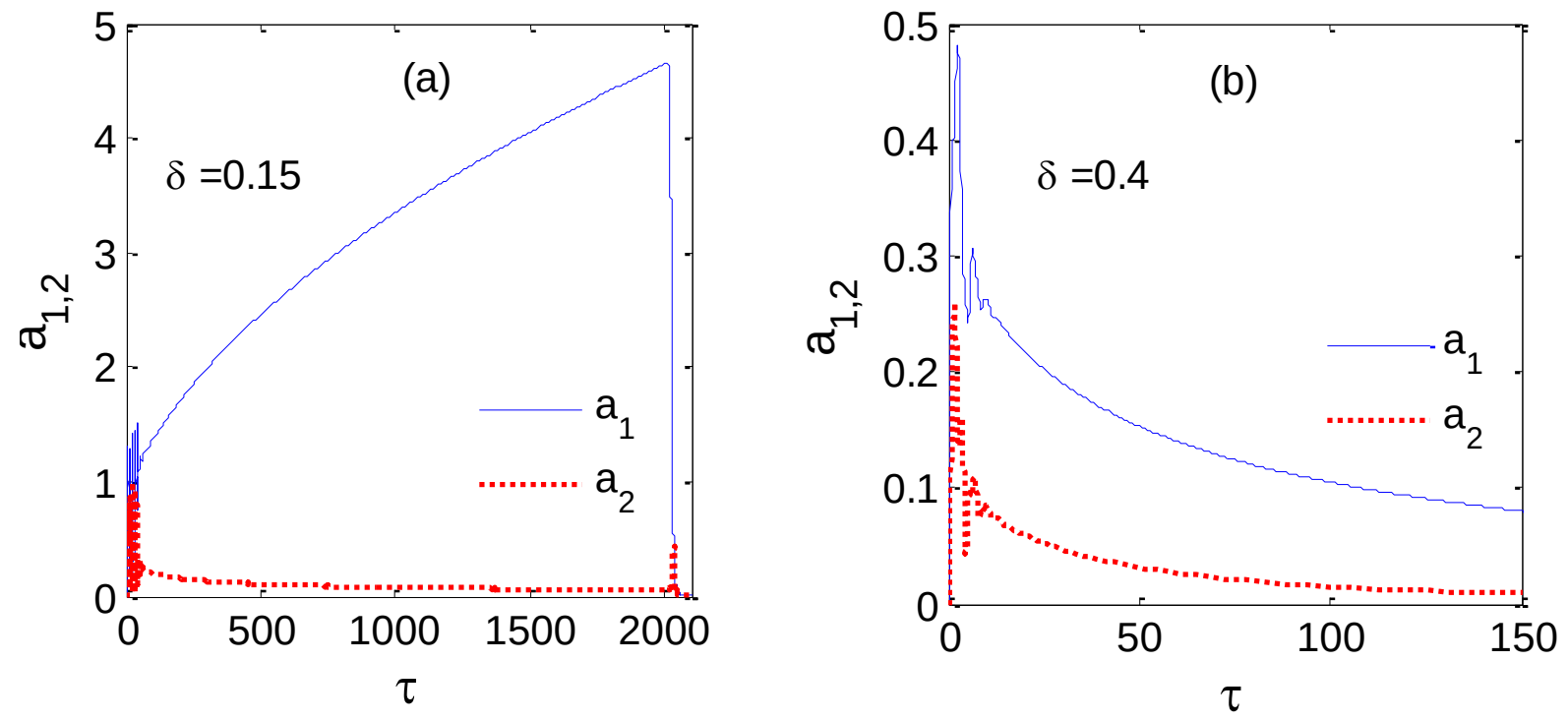

**Fig. 4.** Responses of the coupled oscillators with parameters $\varepsilon = 0.13, f = 0.7, \beta = 0.04$: (*a*) $\delta = 0.15$; (*b*) $\delta = 0.4$.

Figure 5 illustrates the short-time irregular response enhancement with the subsequent anti-phase escape from resonance for both oscillators in the chain with parameters $\varepsilon = 0.13, f = 0.7, \delta = 0.04$ but $\beta = 0.06 > \beta^*$.

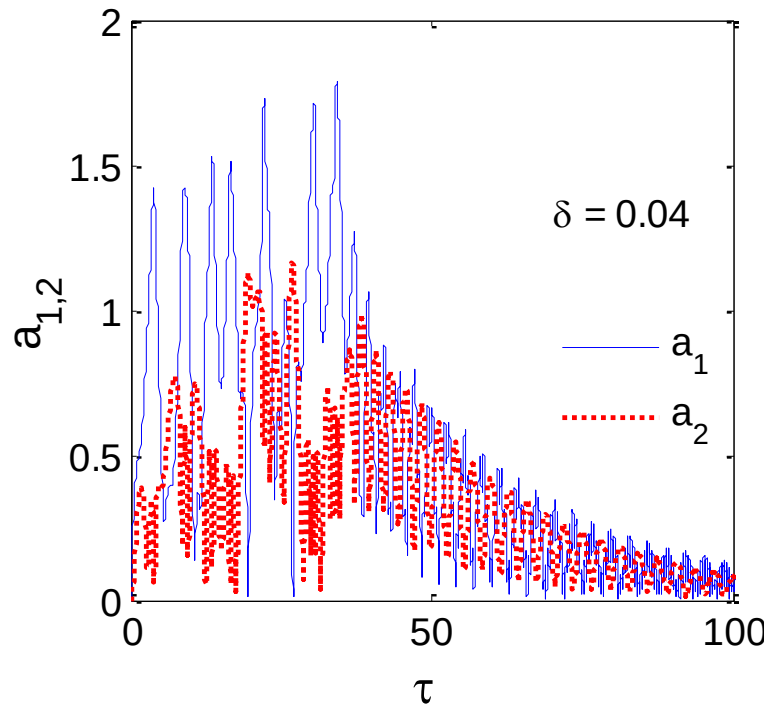

**Fig. 5.** Non-resonant oscillations of the 2-particle array with parameters $\varepsilon = 0.13, f = 0.7, \beta = 0.06 > \beta^*, \delta = 0.04$.

## *4.2. Four-particle arrays*

In this section, we study the dynamics of the 4-particle arrays. First, we consider the chain with parameters $\varepsilon = 0.07, f = 2, \delta = 0.125, \beta = 0.02$. It is easy to conclude that $\delta_*^{(4)} =$ 0.33, and the parameter $\delta = 0.125 < \delta_*^{(4)}$ can be considered as an admissible coefficient of dissipation. Besides, we obtain from (A.5), (A.6) that the threshold for the detuning rate in the undamped chain is equal to $\beta^* \approx 0.72$, that is $\beta << \beta^*$. Under these conditions, the entire chain can be captured into resonance (Fig. 6). Figure 6($a$) illustrates the responses in the initial time interval, wherein the amplitude $a_1(\tau)$ exceeds the other amplitudes oscillating near the common backbone curve (cf. Fig. 3($a$)). Due to similarity of the resonant behavior at large times, Fig 6($b$) depicts only the amplitude of the first oscillator. The simultaneous escape from resonance of all particles is shown in Fig. 6($c$). It follows from Fig. 6($c$) that the approximations $\tau_{AR} \approx 5360$, $\bar{a}(\tau_{AR}) = 4$ are close to the exact (numerical) solutions at the instant of escape from resonance.

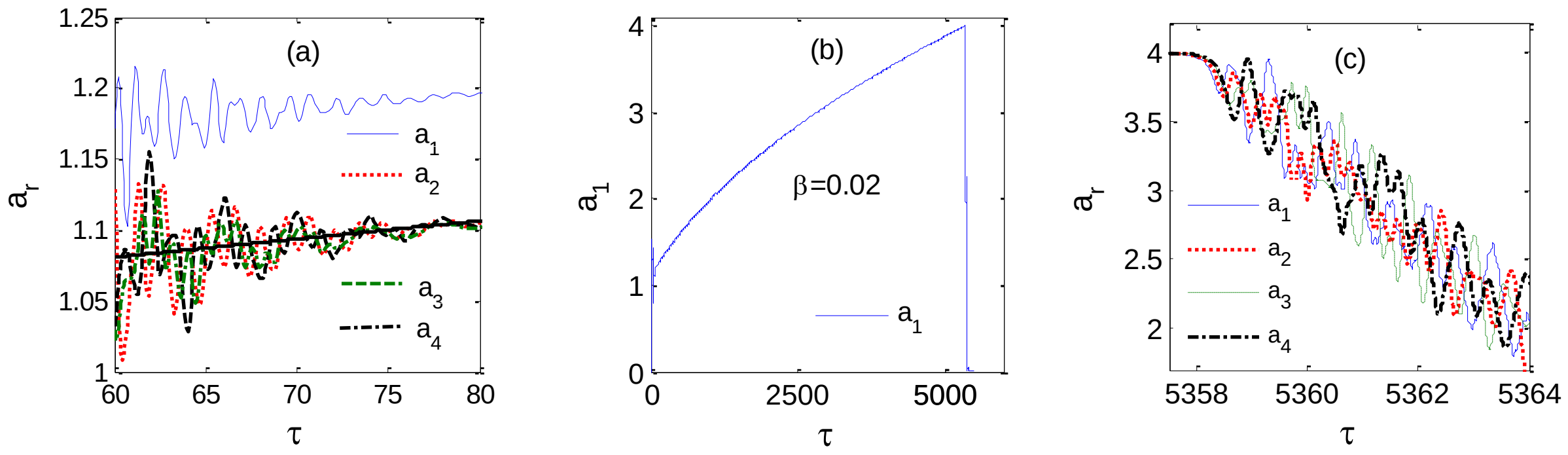

**Fig. 6.** Autoresonance in the coupled oscillators with parameters ($\varepsilon = 0.07$, $f = 2$) $\in D_1$, $\beta = 0.02$, $\delta = 0.125$: ($a$) the amplitudes of all oscillators in the initial interval of motion; the backbone curve $\bar{a}(\tau)$ is indicated by the solid line; ($b$) the amplitude of the first oscillator $a_1(\tau)$ up to escape from resonance capture; ($c$) escape from resonance of all oscillators at $\tau > \tau_{AR}$.

Recall that the threshold $\beta^*$ is established for a single oscillator, and an increase of detuning rate may lead to failure of resonance in the multi-particle array even at $\beta < \beta^*$. Autoresonant energy localization at $\beta = 0.03 << \beta^*$ is illustrated in Fig. 7.

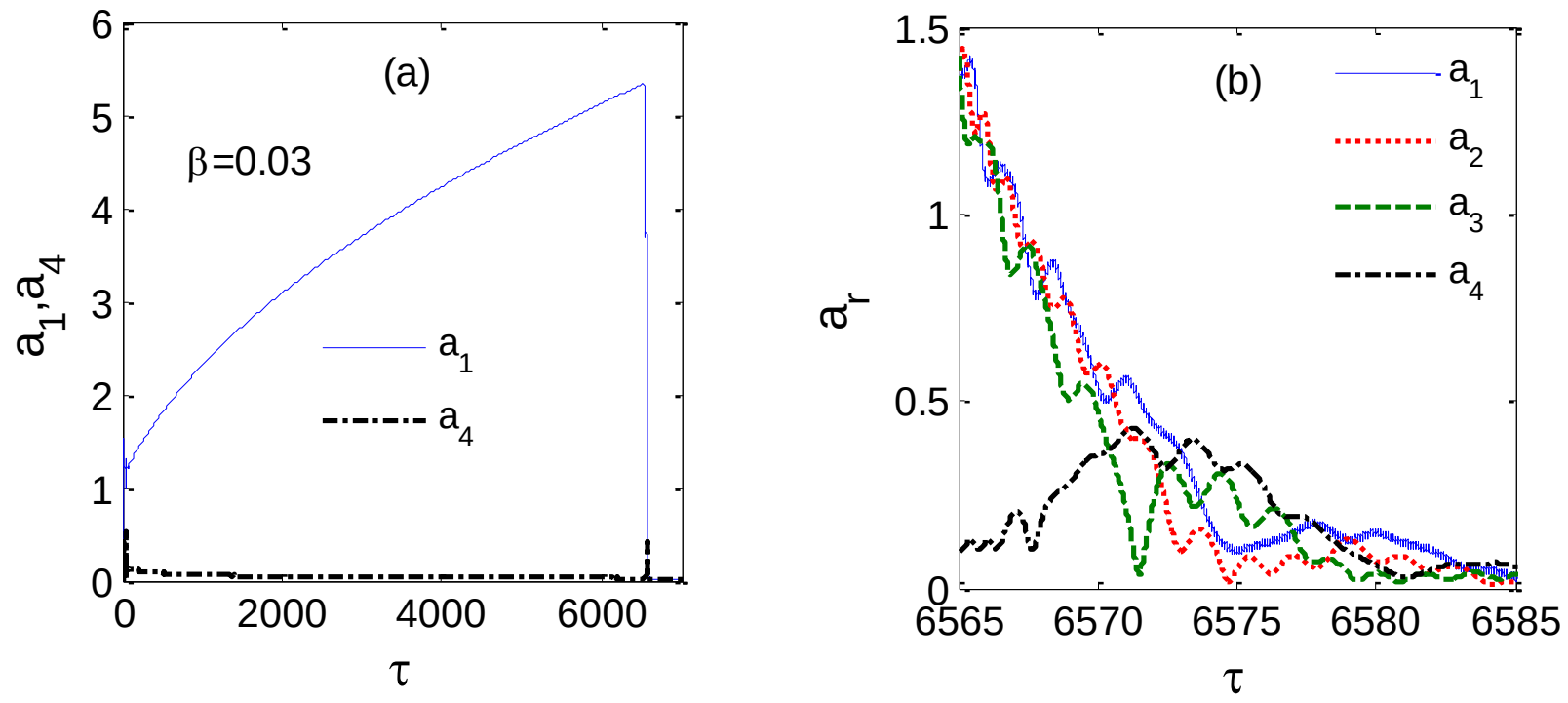

**Fig**. 7. Responses of the coupled oscillators with parameters $\varepsilon = 0.07, f = 2, \beta = 0.03, \delta = 0.125$: (*a*) autoresonance in the first oscillator and non-resonant amplitude of the last oscillator; the resonant amplitudes of the $2^{nd}$ and $3^{rd}$ oscillators are identical to $a_1(\tau)$; (*b*) simultaneous escape from resonance of the autoresonant oscillators and small oscillations of the fourth oscillator.

Figure 7(*a*) depicts autoresonance in the first oscillator and non-resonant oscillations of the fourth oscillator; the amplitudes of the $2^{nd}$ and $3^{rd}$ oscillator are close to $a_1$ (see Fig. 7(*b*)). It follows from (34) that the quasi-steady approximations $\tau_{AR} \approx 6540$, $\bar{a}(\tau_{AR}) = 5.33$, $r \in [1, 3]$ are close to the exact (numerical) solutions at $\tau = \tau_{AR}$.

Figure 8 demonstrates energy localization in the first pair of oscillator at $\beta = 0.04$. Figure 8(*a*) depicts the final interval of the response enhancement for the first oscillator; the autoresonant amplitude $a_2(\tau)$ is identical to $a_1(\tau)$ in the chosen time- and space-scales; the transition from chaotic oscillations to autoresonance is shown in Fig. 8(*b*); small non-resonant oscillations in the last oscillators are illustrated in Fig. 8(*c*). Figure 8 proves that the approximate values of the exit time $\tau_{AR} = 11250$ and the amplitudes $\bar{a}_r(\tau_{AR}) = 8$, $r = 1, 2$, are close to the exact (numerical) solutions at $\tau = \tau_{AR}$.

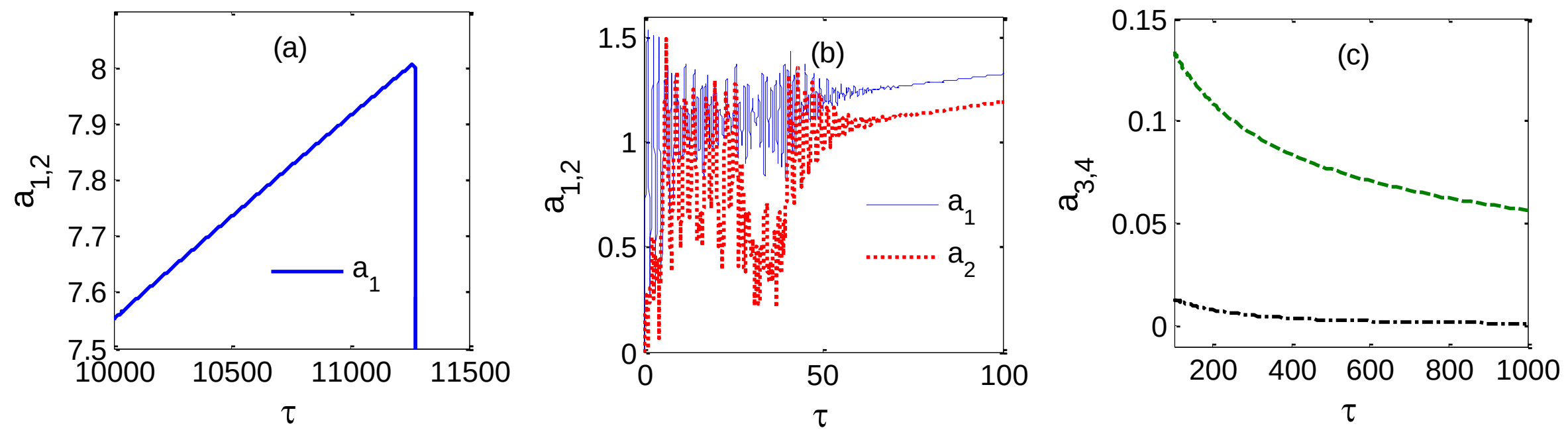

**Fig. 8**. Energy localization in the oscillator chain with parameters $\varepsilon = 0.07$, $f = 2$, $\beta = 0.04$, $\delta = 0.125$: (*a*) response enhancement in the first oscillator; the amplitude $a_2(\tau)$ is similar to $a_1(\tau)$ in the chosen time- and space-scales; (*b*) transitions from chaotic oscillations to autoresonance in the first pair of oscillators; (*c*) small non-resonant oscillations in the last pair oscillators.

The effect of dissipation is illustrated in Fig. 9 for the chain with parameters $\varepsilon = 0.07$, $f = 2$, $\beta = 0.02$ and the parameters of dissipation $\delta = 0.2$ and $\delta = 0.25$. Figures 9(*a*), 9(*b*) depict autoresonance in the first particles, Fig. 9(*c*) illustrates non-resonant dynamics of the last particle at $\delta = 0.2$. As inferred from Fig. 9(*b*), initial irregular oscillations of the autoresonant particles quickly change to motion with monotonically increasing amplitudes. Since the amplitudes $a_r(\tau)$, $r = 1,2,3$, are nearly identical to each other in the chosen time- and scale-space, Fig. 9(*a*) demonstrates autoresonance only in the first oscillator. Autoresonant energy localization in the first pair of oscillators and small non-resonant oscillations in the last pair of oscillators at $\delta = 0.25$ are shown in Figs. 9(*d*), 9(*e*) and 9(*f*), respectively. The quasi-steady approximations $\tau_{AR} \approx 3610$, $\bar{a}(\tau_{AR}) \approx 3.33$ ($r$ = 1, 2, 3) at $\delta = 0.2$, and $\tau_{AR} \approx 5360$, $\bar{a}(\tau_{AR}) = 4$ ($r$ = 1, 2) at $\delta = 0.25$, respectively, agree with the numerical results in Figs. 9(*a*), 9(*d*).

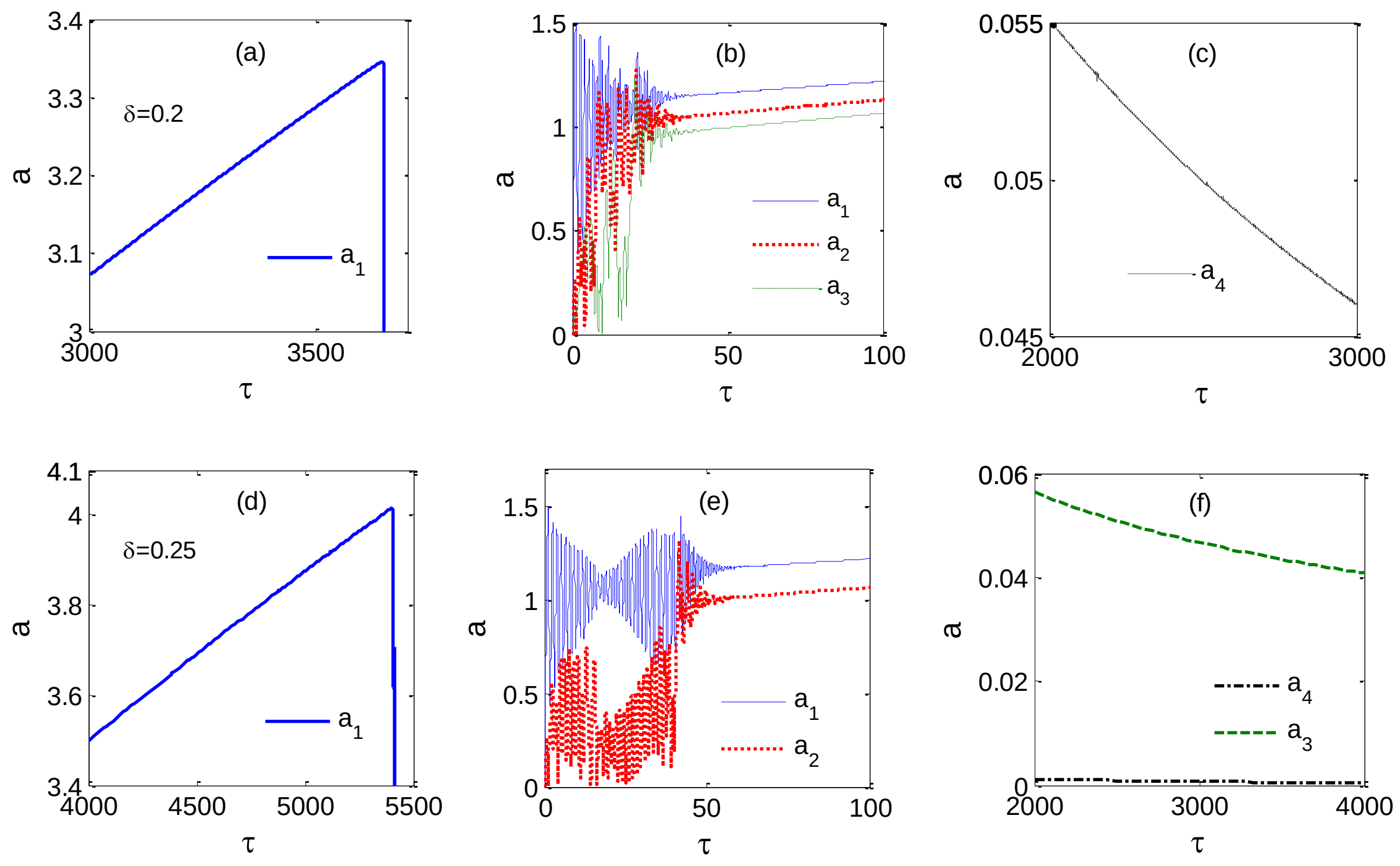

**Fig. 9**. Autoresonance in the chain with parameters $\varepsilon = 0.07$, $f = 2$, $\beta = 0.02$: (*a*) the response amplitude $a_1(\tau)$ at $\delta = 0.2$; (*b*) transitions from chaotic oscillations to autoresonance with similar amplitudes at $r = 1,2,3$, $\delta = 0.2$; (*c*) small non-resonant oscillations in the last oscillator at $\delta = 0.2$; (*d*) autoresonance of the first oscillator at $\delta = 0.25$; (*e*) transitions from chaotic oscillations to autoresonance in the first pair of oscillators at $\delta = 0.25$ in the initial time interval; (*f*) non-resonant oscillations of the last pair of particles at $\delta = 0.25$.

Note that only the two- and four-particle arrays have been considered in this work. Multi-particle arrays can be investigated in a similar way but they become more sensitive to the modification of structural and excitation parameters.

## 5. CONCLUSIONS

In this paper, we have demonstrated that autoresonance can be considered as an effective tool for exciting high energies in nonlinear damped oscillators due to properly chosen slow variations of the resonant frequency. In order to elucidate the system behavior, we have studied analytically and numerically the emergence and stability of autoresonance in a nonlinear weakly damped chain driven by harmonic force with a slowly time-varying frequency. Using the asymptotic procedures, we have constructed the averaged equations, which describe the dynamics of the particles under the condition of 1:1 (fundamental) resonance. It has been proved that, in contrast to unbounded autoresonance in a non-dissipative array, the energy of a weakly damped array is growing only in a bounded time interval up to an instant of escape from resonance. The amplitudes of autoresonant oscillations at large times can be approximately described as the superposition of small fluctuations to the monotonically increasing backbone curve found from the quasi-steady equations and common for all autoresonant oscillators. This leads to formal energy equipartition between the autoresonant oscillators at large times, although the energy of the excited oscillator exceeds the energy of the attachment in the initial time interval. However, energy equipartition does not entail the equal phases, which explicitly depend on the amplitude of external forcing.

The derived equations yield both the duration of autoresonance and the attainable amplitudes and phases. Energy localization in the autoresonant part of the chain with the approximate energy equipartition between the oscillators is described both analytically and numerically. Close proximity of the analytical and numerical results has been demonstrated for several multi-particle arrays. To the best of the author's knowledge, these results have not been obtained in earlier works.

## APPENDIX

This Appendix provides a concise presentation of the conditions of autoresonance for the non-dissipative single oscillator. Additional details can be found in [15].

It was shown [12] that, under assumptions of Sec. III, the solution $a_1(\tau)$ of the basic single oscillator (14) is very close to the solution of a similar time-independent oscillator at sufficiently small time $\tau$. Thus the first step towards analyzing autoresonance is the study of the transition from small to large oscillations in the underlying non-dissipative system with the constant excitation frequency. Equations (14) at $\delta = 0$, $\zeta_0 = 0$ take the form:

$$\frac{da_1}{d\tau} = -f \sin \Delta_1, \tag{A.1}$$

$$a_1 \frac{d\Delta_1}{d\tau} = \sigma(a_1^2 - 1)a_1 + a_1 - f \cos \Delta_1$$

with initial condition $a_1(0) = 0$, $\Delta = -\pi/2$. It was proved [12] that the transition from small to large oscillations in the oscillator (A.1) occurs due to the loss of stability of the LPT of small oscillations at a critical value $f = f_{1\varepsilon}$ of the forcing amplitude. The amplitude $f$ corresponding to large oscillations is defined as (see [28](21)):

$$f > f_{1\varepsilon} = \sqrt{(1 - 2\varepsilon)^3/54\varepsilon^2}. \tag{A.2}$$

Note that $d(f_{1\varepsilon})^2/d\varepsilon \to -1/(27\varepsilon^3)$ as $\varepsilon \to 0$. This implies that a decrease of the coupling response $\varepsilon$ entails the growth of the threshold $f_{1\varepsilon}$.

The next step is to define the admissible values of the parameter $\varepsilon$, which yield the coupling parameter $\varepsilon$ sufficient to sustain resonance in the *r-th* oscillator under the condition of resonance in the previous oscillator and small oscillations of the subsequent oscillator. It was shown [15] that the admissible values of the coupling strength $\varepsilon$ are defined by the condition:

$$\varepsilon > \varepsilon_{cr} = 0.125. \tag{A.3}$$

Conditions (A.2), (A.3) are presented in Fig. 10.

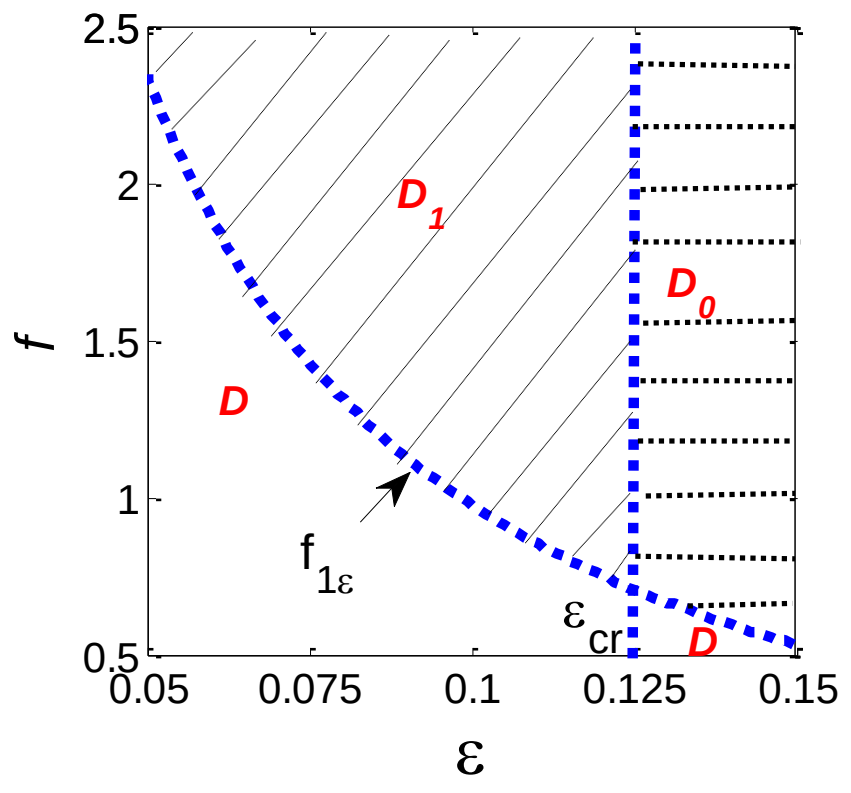

**Fig. 10**. Parametric boundaries (A.2), (A.3) for the oscillator (A.1): the entire chain with the parameters $(\varepsilon, f) \in D_0$ is captured into resonance; all particles with the parameters $(\varepsilon, f) \in D$ execute small oscillations; if $(\varepsilon, f) \in D_1$, then the forced oscillator is captured into resonance but the dynamics of the attachment should be investigated separately.

We can conclude that the non-dissipative chain with parameters $(\varepsilon, f) \in D$ is non-resonant but the chain with parameters $(\varepsilon, f) \in D_0$ is entirely captured into resonance; if $(\varepsilon, f) \in D_1$, then the forced oscillator is resonant but the dynamics of the attachment should be investigated separately.

Now we consider the non-dissipative analogue of the autoresonant oscillator (14), namely,

$$\frac{da_1}{d\tau} = -f \sin \varDelta_1, \tag{A.4}$$

$$a_1 \frac{d\varDelta_1}{d\tau} = \sigma(a_1^2 - 1)a_1 - \zeta_0(\tau)a_1 + a_1 - f \cos \varDelta_1,$$

with initial conditions $a_1(0) = 0$, $\varDelta = -\pi/2$. Note that the emergence of autoresonance in the undamped oscillator also depends on the critical detuning rate $\beta^*$, at which the transition from bounded to unbounded response takes place. The response amplitudes of oscillator (A.4) with parameters $(\varepsilon = 0.13,\ f = 0.7) \in D_0$ and detuning rate $\beta$ are presented in Fig. 11. For comparison, Fig. 11 also depicts the response amplitude of the time-invariant oscillator with the "frozen" coefficient $\zeta_0^* = \beta T^*$, where $T^*$ is an instant of inflection for the LPT of the basic oscillator described by the following equations

$$\frac{da_1}{d\tau} = -f\sin\Delta_1, \tag{A.5}$$

$$a_1\frac{d\Delta_1}{d\tau} = \sigma(a_1^2 - 1)a_1 - \zeta_0^*\, a_1 + a_1 - f\cos\Delta_1.$$

It is seen in Fig. 11 that the LPT of the oscillator (A.4) has a noticeable inflection at $\tau = T^*$, and the transitions from small to large oscillations in the adiabatic system (A.4) also takes place at $\tau \approx T^*$, despite the divergence of the solutions at $\tau > T^*$.

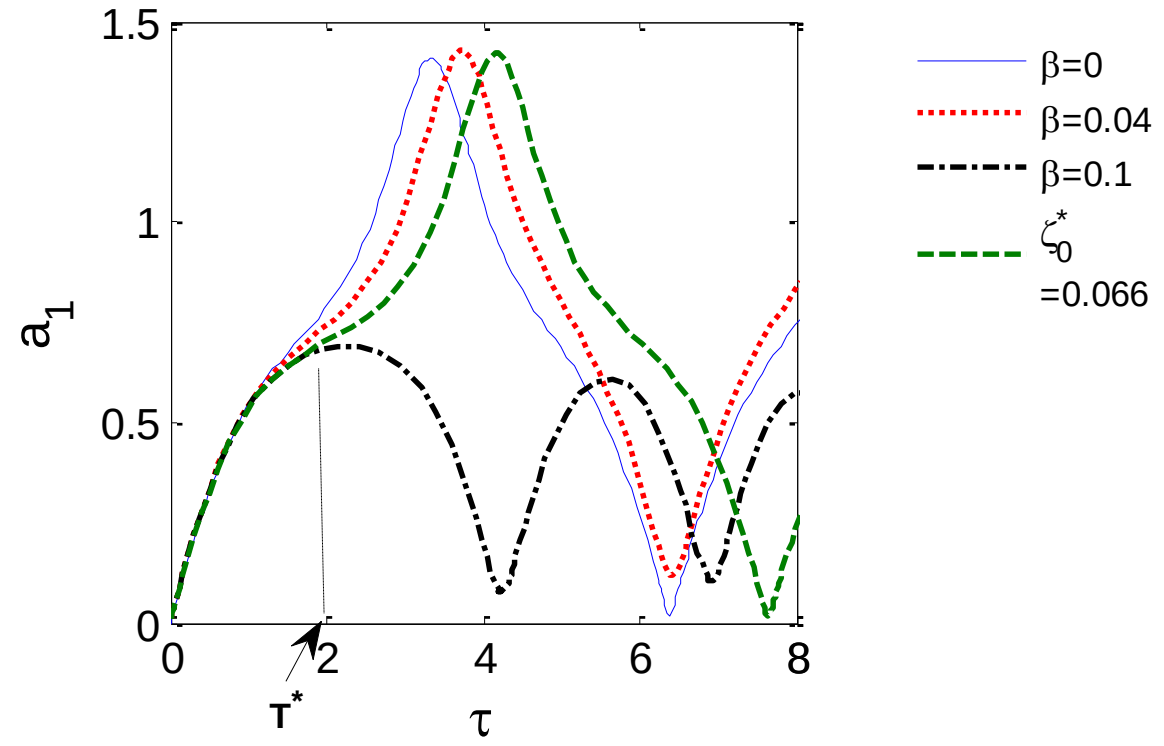

**Fig. 11**. Response amplitudes of the oscillators (A.1), (A.4), (A.5) with the parameters ($\varepsilon = 0.13, f = 0.7$) $\in D_0$ and different detuning rates $\beta$. The inflection point $T^* \approx 1.65$ for the oscillator (A.1) (the solid line) is close to the inflection point for the adiabatic system (A.4) at $\beta = 0.04$ (the dotted line).

From Fig. 11, it is seen that the response amplitude of the oscillator (A.4) (the dotted line) lies between the LPTs of the time-invariant oscillator (A.1) (the solid line) and the oscillator (A.5) with the "frozen" parameter $\zeta_0^*$ (the dashed line). This implies that capture into resonance of the model (A.5) with the "frozen" detuning may be considered as a sufficient condition of the emergence of AR in the adiabatic system (A.4).

Considering $\zeta_0(T^*) = \beta T^*$ as a "frozen" parameter and using the results from [15], we express the following condition of the emergence of autoresonance in the adiabatic oscillator:

$$2\varepsilon\beta T^* < (1 - 2\varepsilon)[(f/f_{1\varepsilon})^{\frac{2}{3}} - 1], \tag{A.6}$$

where $T^*$ denotes the instant of inflection for the LPT of the oscillator (A.1) with the same parameters $\varepsilon$ and $f$ (e.g., see [15]). Under this assumption, it follows from (A.5) that critical detuning rate $\beta^*$ is given by

$$\beta^* = (1 - 2\varepsilon)\frac{[(f/f_{1\varepsilon})^{\frac{2}{3}} - 1]}{2\varepsilon T^*}, \tag{A.7}$$

and $d\beta^*/d\varepsilon > 0$ at $\varepsilon < 1/(\sqrt{2}f)$. This means that critical rate $\beta^*$ increases with increasing coupling strength $\varepsilon$ under a fixed forcing amplitude $f > f_{1\varepsilon}$.

The condition $\beta < \beta^*$ admits the emergence of autoresonance in the oscillator (A.1). To improve the correctness of numerical results, in practical problems it is convenient to employ the numerically found values of the inflection point $T^*$. In particular, it was found [15] that $T^* \approx 1.65$, $\beta^* \approx 0.058$ at $\varepsilon = 0.13$, $f = 0.7$ (see Fig. 11).